\pdfoutput=1

\documentclass[nofootinbib,twocolumn,showpacs,aps,prd,superscriptaddress,amsmath,amssymb,floatfix,a4paper,twoside]{revtex4-1}
\usepackage{graphicx}
\usepackage{bm}
\usepackage{color}
\usepackage{multirow}
\usepackage{bookmark}

\begin{document}

\title{Radiation of a relativistic electron with non-equilibrium own Coulomb field}

\author{A.S.~Fomin}
\affiliation{NSC Kharkiv Institute of Physics and Technology, 61108 Kharkiv, Ukraine}
\author{S.P.~Fomin}
\email[]{sfomin@kipt.kharkov.ua}
\affiliation{NSC Kharkiv Institute of Physics and Technology, 61108 Kharkiv, Ukraine}
\author{N.F.~Shul'ga}
\affiliation{NSC Kharkiv Institute of Physics and Technology, 61108 Kharkiv, Ukraine}

\date{December 22, 2010}

\begin{abstract}

The condition and specific features of non-dipole regime of radiation is discussed in connection with the results of recent CERN experiment NA63 on measurement of radiation power spectrum of $149\,$GeV electrons in thin tantalum targets.
The first experimental detection of logarithmic dependence of radiation yield from the target thickness is the conclusive evidence of the effect of radiation suppression in a thin layer of matter, which was
predicted many years ago, and which is the direct manifestation of radiation of relativistic electron with non-equilibrium own Coulomb field.
The special features of angular distribution of radiation and its polarization in a thin target at non-dipole regime are proposed for a new experimental study.

\end{abstract}

\pacs{41.60.-m; 41.75.Ht}

\maketitle

\setcounter{footnote}{0}

%
\section{\label{sec:introduction}Introduction}

During last two years there were published the results of recent experimental investigation of special features of relativistic electron bremsstrahlung in a thin target
\cite{Thomsen:2009,Thomsen:2010}.
These measurements were done by international collaboration NA63 in CERN using SPS secondary electron beam with energy around $200\,$GeV.
One of the main motives to carry out this experiment was to clear the unexpected result of the SLAC experiment E-146
\cite{Klein:1993,Anthony,Klein:1999} that showed a strange behavior of the radiation spectrum of $25\,$GeV electrons in relatively small thickness target, especially for the gold target of $0.7\,\%$ of radiation length thickness 
\cite{Klein:1993,Anthony}.

The SLAC experiment E-146 was generally devoted to the verification of the Migdal quantitative theory of the Landau-Pomeranchuk-Migdal (LPM) effect
\cite{Landau,Migdal},
which describes the suppression of radiation of relativistic electrons in an amorphous matter due to the multiple scattering on atoms in comparison with the predictions of the Bethe-Heitler theory
\cite{Bethe-Heitler}.
The analysis of the data obtained in SLAC experiment E-146 showed a good agreement between the calculations using the Migdal formula (LPM effect) and the experimental data for relatively thick targets and not very low photon energies.
However, for the case of golden target of $0.7\,\%$ of radiation length thickness there was a significant disagreement between theory and experiment 
\cite{Anthony}.
Such ``unexpected" behavior of the radiation spectrum at low frequencies was named in
\cite{Anthony}
as ``edge effect" and firstly they tried to exclude it by subtraction procedure, because ``no satisfactory theoretical treatment of this phenomenon" was found for that moment. Actually, they sought out the Ternovskii article 
\cite{Ternovskii},
in which the Migdal theory of the LPM effect developed for boundless amorphous medium was improved for the finite target thickness case.
However, when they tried to use the Ternovskii formula to describe the ``edge effect", they obtained the excess of the Bethe-Heitler result
\cite{Bethe-Heitler}
instead of the expected suppression, and they wrote in 
\cite{Klein:1993}
that this formula gives ``unphysical result".

The observed in SLAC experiment discrepancy stimulated a new wave of theoretical investigations of the multiple scattering effect on radiation (see
\cite{Shul'ga:1998:Lett,Shul'ga:1998,Blankenbacler,Zhakharov,Baier:1996,Baier:1999}).
In
\cite{Shul'ga:1998:Lett}
it was shown that observed in
\cite{Klein:1993,Anthony,Klein:1999}
deviation from predictions of the Migdal theory takes place, when the target thickness $t$ was small in comparison with a coherence length (or formation zone) of radiation process
$l_{\rm c}=2e^{\prime}e/m^{2}\omega$
\cite{Ter-Mikaelian}
(here $m$ and $e$ are the mass and initial energy of an electron, $\omega$  is the emitted photon energy, $e^{\prime}=e-\omega$, we use the system of units: $h=c=1$).
Exactly this case $t\ll{l}_{\rm c}$ was considered earlier theoretically in
\cite{Shul'ga:1978,Fomin:1986},
where a specific effect of the suppression of radiation in a thin layer of matter was described and discussed in details including its essential distinction from the LPM and BH regimes of radiation.
As it was shown in 
\cite{Shul'ga:1998:Lett},
the ``unphysical result" obtained by the Ternovskii formula in
\cite{Klein:1993} was connected with the usage of the asymptotic formula for a mean-square angle of multiple scattering, which is not applicable for the SLAC experiment E-146 conditions.

The quantitative theory of the radiation suppression effect in a thin layer of matter was developed later in
\cite{Shul'ga:1998:Lett,Shul'ga:1998,Blankenbacler,Zhakharov,Baier:1996,Baier:1999}
using different approaches.
The results obtained in these works are in a good agreement with the SLAC experimental data for the thin golden target (see, for example, reviews
\cite{Anthony,Akhiezer}).
However, it was the only one explicit manifestation of this effect during the SLAC experiment E-146 and it took place in a relatively narrow photon energy region for $25\,$GeV electrons.
That is why it was necessary to carry out a special experimental investigation of this effect at higher electron energy that gives a wider photon energy region for observation of this effect and, what is more important, to study the thickness dependence of radiation intensity for a thin target case, which is fundamentally differed from the BH and LPM regime of radiation (see 
\cite{Shul'ga:1978,Fomin:1986,Akhiezer}).

A new experimental study of the LPM and analogous effects at essentially higher electron energies (up to $\varepsilon=287\,$GeV) was carried out recently at CERN by collaboration NA63 (see 
\cite{Hansen, Uggerhoj}
and also
\cite{Thomsen:2009,Thomsen:2010}).
The results of measurements
\cite{Hansen}
for Ir, Ta and Cu targets of thicknesses about $4\,\%$ of radiation length showed a good agreement with the Migdal theory of the LPM effect.
The effect of suppression of radiation in a thin target, named in
\cite{Hansen}
as the Ternovskii-Shul'ga-Fomin (TSF) effect, was also considered in
\cite{Hansen, Uggerhoj},
however, the photon energy region, in which the TSF effect could be observed for chosen target thicknesses, were below the energy threshold of measured photons $\omega_{\rm min}=2\,$GeV for both these experiments.
The condition for the successful observation of the TSF effect in radiation spectrum was realized later in CERN for 206 and $234\,$GeV electrons radiation in 5--10$\,\mu$m thick Ta targets
\cite{Thomsen:2009}.

Finally, probably the most complicated measurements for realization, but the most important for demonstration of the TSF effect essence, namely the logarithmic thickness dependence of radiation intensity in a thin target, were successfully carried out recently by the CERN collaboration NA63 
\cite{Thomsen:2010}.
This is the first direct demonstration of the suppression of radiation effect for a relativistic electron with non-equilibrium own Coulomb field 
\cite{Fomin:1986,Feinberg}.
This effect should have its analog also in QCD at quark-gluon interaction.

In this paper we present the theoretical analysis and treatment of the recent CERN experimental results 
\cite{Thomsen:2009,Thomsen:2010}.
We also propose to carry out a new experiment to study the special features of the angular distribution of radiation at the TSF effect conditions, which were theoretically described in
\cite{Fomin:2003}.
These features can give a new opportunity for obtaining a high degree of linear polarization of gamma-quanta that was proposed in
\cite{Fomin:2005}.

%
\section{\label{sec:conditions}General conditions and features of LPM and TSF effects}

According to the standard Bethe-Hietler theory of bresstrahlung in amorphous matter the radiation power spectrum $dE/d\omega$ defined by scattering of relativistic electron on target atoms is proportional to the target thickness $t$
\cite{Bethe-Heitler}
\begin{equation}\label{eq:BH}
    \frac{dE_{\rm BH}}{d\omega}=\frac{2t}{3X_{0}}\left[\left(1+
    \frac{{\varepsilon^{\prime}}^{2}}{\varepsilon^{2}}\right)+
    \frac{\omega^{2}}{2\varepsilon^{2}}\right]\,,
\end{equation}
where $X_{0}$ is the radiation length of target material.

Landau and Pomeranchuk showed
\cite{Landau}
that if the root-mean-square angle of electron multiple scattering $\theta_{\rm ms}$ at the distance of a coherence length $l_{\rm c}$ exceeds the characteristic angle of relativistic particle radiation $\theta\sim\gamma^{-1}$, where $\gamma=\varepsilon/m$ is the Lorentz factor of an electron, then the radiation power spectrum will be suppressed in comparison with
the Bethe-Hietler result given by formula~(\ref{eq:BH}).

The root-mean-square angle of electron multiple scattering on atoms in an amorphous medium at the depth $t$ is inversely proportional to the electron energy $\varepsilon$
\cite{Bethe-Heitler,Akhiezer}
\begin{eqnarray}\label{eq:Qms}
    &&\theta_{\rm ms}(t)=(\varepsilon_{\rm s}/\varepsilon)\sqrt{t/X_{0}}\left[1+0.038\ ln(t/X_{0})\right]\,,
    \nonumber \\
    &&\varepsilon_{\rm s}^{2}=4\pi\cdot{m}^{2}/e^{2}\,,
\end{eqnarray}
so, the target thickness $l_{\gamma}$, at which
$\theta_{ms}(l_{\gamma})=\gamma^{-1}$ does not depend on the electron energy $\varepsilon$ and is determined by the target material only $l_{\gamma}\approx 0.15\,\%\,X_{0}$.

Thus, the condition of the suppression of radiation due to the multiple scattering effect $\theta_{\rm ms}(l_{\rm c})>\gamma^{-1}$ (so-called non-dipole regime of radiation) can be written in the following form:
\begin{equation}\label{eq:nondipole}
    l_{\rm c}>l_(\gamma)\,.
\end{equation}

If $t<l_{\gamma}$, i.e. the target thickness $t$ is less than $0.15\,\%\,X_{0}$, the spectral density of radiation for all possible emitted photon energies is defined by the Bethe-Heitler formula~(\ref{eq:BH}).

If $t>l_{\gamma}$, there are three possible regimes of radiation in this case depending on the energy region of emitted photon.

For the relatively hard part of emitted spectrum, when $l_{\rm c}<l_{\gamma}$, we have a dipole regime of radiation describing by the Bethe-Hietler formula~(\ref{eq:BH})
too.

For the non-dipole radiation, $l_{\rm c}>l_{\gamma}$, there are two regions, defined by the ratio between the coherence length $l_{\rm c}$ and the target thickness $t$, with quite different behavior of radiation spectrum. If the target thick enough, $t\gg{l}_{\rm c}>l_{\gamma}$, the Migdal theory
\cite{Migdal}
of the LPM effect, which describes the suppression of radiation in a boundless amorphous medium, is applicable. For relatively thin target, $l_{\rm c}\gg{t}>l_{\gamma}$ (intermediate case), the TSF mechanism of radiation
\cite{Ternovskii,Shul'ga:1978}
is realized.

The condition~(\ref{eq:nondipole})
determines the photon energy region, where the LPM effect is essential:
\begin{eqnarray}\label{eq:LPMcondition}
    &&\omega<\omega_{\rm LPM}=\frac{\varepsilon}{1+\varepsilon_{\rm LPM}/\varepsilon}\,,\\
    &&\varepsilon_{\rm LPM}=\frac{e^{2}m^{2}}{4\pi}X_{0}\approx7.7\,{\rm TeV}\cdot{X}_{0}\,({\rm cm})\,,
    \nonumber
\end{eqnarray}
It means that for ultra high electron energy $(\varepsilon\gg\varepsilon_{\rm LPM})$ the whole radiation spectrum is suppressed due to the LPM effect: $\omega_{\rm LPM}\approx\varepsilon$.

If $\varepsilon\ll\varepsilon_{\rm LPM}$, then \mbox{$\omega_{\rm LPM}\approx\varepsilon^{2}/\varepsilon_{\rm LPM}\approx1600\,\gamma^{2}/X_{0}.$}

The Migdal function $\Phi_{\rm M}$
\cite{Migdal}
describes the deviation of the radiation spectrum for $\omega<\omega_{\rm LPM}$ from the Beth-Hietler formula~(\ref{eq:BH})
in relatively soft part of the spectrum $(\omega\ll\varepsilon)$:
\begin{eqnarray}\label{eq:LPMspectra}
    &&\frac{dE_{\rm LPM}}{d\omega}\approx\frac{dE_{\rm BH}}{d\omega}\Phi_{\rm M}(s)\,,\\
    &&\Phi_{\rm M}(s)=24s^{2}\left(\int_{0}^{\infty}{dx\ {\rm ctg}(x)\,e^{-2sx}-\frac{\pi}{4}}\right)\,, \nonumber\\
    &&s=\frac{1}{2}\sqrt{\frac{\omega}{2\, \omega_{\rm LPM}}}\,. \nonumber
\end{eqnarray}

The upper limit for the emitted photon energy for the TSF regime of radiation follows from the TSF effect condition
\begin{equation}\label{eq:TSFcondition}
    l_{\rm c}\gg{t}>l_{\gamma}\,.
\end{equation}
It is defined by equality $t = l_{\rm c}$ and can be written in the following form:
\begin{eqnarray}\label{eq:TSFcondition2}
    &&\omega_{\rm TSF}=\frac{\varepsilon}{1+\varepsilon_{\rm TSF}/\varepsilon}\,,\\
    &&\varepsilon_{\rm TSF}=\frac{m^{2}t}{2}\approx6.6\,{\rm PeV}\cdot{t}({\rm cm})\,.
    \nonumber
\end{eqnarray}

If $\varepsilon\gg\varepsilon_{\rm TSF}$, one can use simpler expression for the TSF effect threshold $\omega_{\rm TSF}\approx2\gamma^{2}/t$.

The quantitative theory of the multiple scattering effect on radiation of relativistic electron in a thin layer of matter (the TSF effect) was developed in
\cite{Shul'ga:1998:Lett}
using classical formulas for spectral density of radiation and the results of the Bethe-Moliere theory of multiple scattering 
\cite{Bethe}.
This approach is valid if $\omega\ll\varepsilon$. Namely such a condition was realized for both experimental investigations at SLAC E-146
\cite{Klein:1993,Anthony}
and at CERN NA63
\cite{Thomsen:2009,Thomsen:2010}.
The radiation power spectrum in this case $(l_{\rm c}\gg{t})$ is determined by formula
\cite{Shul'ga:1998}
\begin{eqnarray}\label{eq:TSFspectra}
   && \frac{dE_{\rm TSF}}{d\omega}=\frac{2e^{2}}{\pi}\int{d\theta_{\rm s}{f}_{\rm BM}(\theta_{\rm s})}
    \left[\frac{2\xi^{2}+1}{\xi\sqrt{\xi^{2}+1}}\,{\rm ln}\left(\xi+\right.\right. \nonumber \\
   && \left.\left.+\sqrt{\xi^{2}+1}\right)-1\right]\,,\qquad
    \xi=\gamma\theta_{\rm s}/2\,,
\end{eqnarray}
in which the averaging over electron multiple scattering in a target carried out with the Bethe-Moliere distribution function $f_{\rm BM}$
\cite{Bethe}
(for details see
\cite{Shul'ga:1998:Lett,Akhiezer}).
At $t\ll{l}_{\gamma}$ (that means $\xi\ll1$) the formula~(\ref{eq:TSFspectra})
gives the Bethe-Hietler result with the linear dependence from the target thickness. In the opposite case, i.e. at $t\gg{l}_{\gamma}$, the formula~(\ref{eq:TSFspectra})
gives only the logarithmic increasing of radiation power spectral density with the target thickness increasing.

Such a strange behavior of radiation power (the scattering angle still increase linearly with thickness, but radiation does not) can be explain by relativistic delay effect during regeneration of the own Coulomb field of relativistic electron after its scattering on a big angle $\theta_{\rm s}>\gamma^{-1}$, and it can be treated as a radiation of the ``half-bare" electron, i.e. the electron with non-equilibrium own Coulomb field (see
\cite{Fomin:1986,Akhiezer,Feinberg}
for detailed discussion). This logarithmic behavior will be replaced by the linear one again when the target thickness reaches the value of coherence length for the given photon energy $\omega$.

The quantum treatment of the TSF effect was done in 
\cite{Shul'ga:1998,Blankenbacler,Zhakharov,Baier:1996,Baier:1999}
using different approaches and it is became an important for ultra high electron energy $\varepsilon\gg\varepsilon_{\rm TSF}$), when $\omega_{\rm TSF}\approx\varepsilon$ and the whole radiation spectrum is suppressed due to the TSF effect.

There are two additional factors that have essential influence on
radiation process in matter, namely, the dielectric suppression (or
the Ter-Mikaelyan effect
\cite{Ter-Mikaelian})
 and the transition radiation from the target bounds
\cite{Klein:1999,Ter-Mikaelian}.
Both these effects could be neglected, if we consider the photons energy higher than $\omega_{0}=\gamma\omega_{\rm p}$ , where $\omega_{\rm p}$ is the plasma frequency
\cite{Ter-Mikaelian}.
For tantalum target and the electron beam energy $\varepsilon=150\,$GeV this threshold is about $\omega_{0}\approx25\,$MeV.

\begin{figure}[tbh]
\begin{center}
\includegraphics[width=0.53\textwidth] {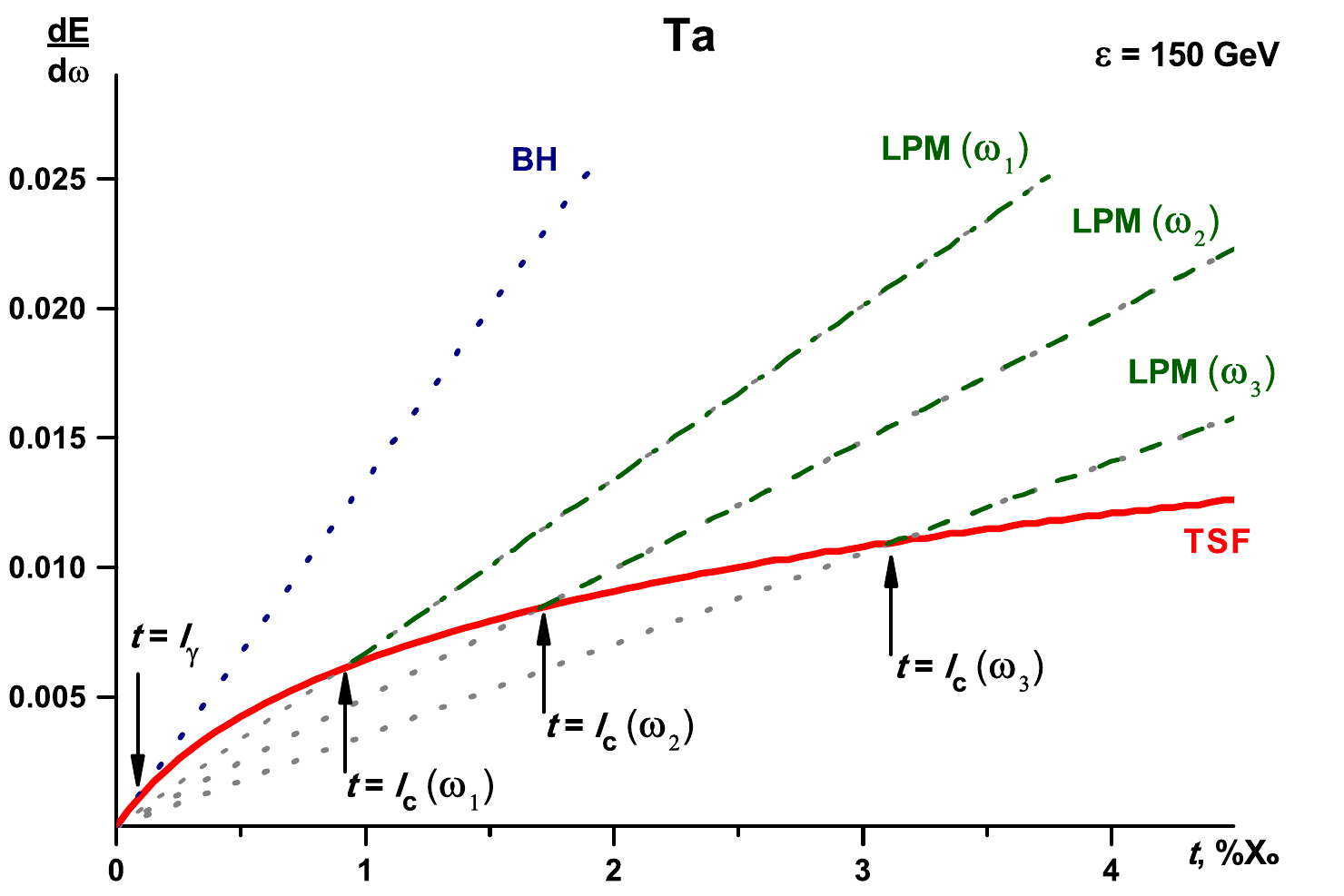}
\end{center}
\caption{The radiation power spectrum of $150\,$GeV electrons in tantalum target via target thickness $t$ $(\%\,X_{0})$. The detailed description of curves is given in text}
\label{fig:spectrum}
\end{figure}

The thickness dependence of the radiation power spectrum $dE/d\omega$ demonstrates clearly the qualitative difference between the different regimes of radiation in amorphous matter, namely the BH, LPM and TSF regimes and their consequent replacements.
The results of theoretical calculations of such dependence are represented in Fig.~\ref{fig:spectrum}.

%
\onecolumngrid

\newpage
\twocolumngrid
%

For $t<l_{\gamma}$, i.e$.$ when the target thickness $t$ is too small that the multiple scattering of relativistic electrons in target is not enough to fulfill the condition (3), the radiation process has a dipole character and the radiation power spectrum is described by the Bethe-Hietler formula~(\ref{eq:BH}).
The soft part of the Bethe-Hietler spectrum $(\omega\ll\varepsilon)$ does not depend on $\omega$ and describes by a very simple formula $dE_{\rm BH}/d\omega=4t/3X_{0}$.
The corresponding curve is presented in Fig.~\ref{fig:spectrum}
by dashed straight line ``BH".

At the target thickness increasing the condition $t=l_{\gamma}$ could be fulfilled, and at this point the dipole regime of radiation is changed for the non-dipole that leads to suppression of radiation comparing with the Bethe-Hietler formula predictions.
For relatively soft photons, for which $l_{\rm c}\gg{t}$, the radiation power for this part of radiation spectrum is determined by the Eq.~(\ref{eq:TSFspectra})
that means the TSF regime of radiation with a logarithmic dependence on the target thickness $t$ (solid line ``TSF" in Fig.~\ref{fig:spectrum}.
As it follows from Eq.~(\ref{eq:TSFspectra})
$dE_{\rm TSF}/d\omega$ does not depend on the emitted photon energy $\omega$, however, the validity condition of the TSF regime (\ref{eq:TSFcondition})
does.
It means that for different $\omega_{n}$ the transition from the TSF to LPM regime of radiation takes place at different values of target thickness $t_{n}=l_{\rm c}(w_{n})$.
In Fig.~\ref{fig:spectrum}
there are three such points marked by arrows for different photon energies $\omega_{n}$, namely $\omega_{1}=150$, $\omega_{2}=350$ and $\omega_{3}=800\,$MeV.
There are also three different dot-dashed lines ``LPM", which are calculated using the Migdal formula~(\ref{eq:LPMspectra})
for these values of photon energy respectively.
Thus, changing the target thickness one can consequently observe three different mechanisms of radiation of relativistic electron in amorphous target such as the BH, TSF and LPM.

The first experimental investigation of the thickness dependence transformation from the linear regime (BH) via logarithmic one (TSF) to the linear (LPM) again was recently done in CERN by collaboration NA63
\cite{Thomsen:2010}.
In spite of all difficulties connected with a very complicate experimental installation and operating with a set of thinnest targets of several microns thickness, this experiment gave a conclusive proof of the suppression effect of relativistic electron radiation in a thin layer of matter predicted many years ago
\cite{Ternovskii,Shul'ga:1978}
and $per$ $se$ it gave the unique demonstration of the space-time evolution of the radiation process in matter by example of relativistic electron with non-equilibrium own Coulomb field 
\cite{Fomin:1986,Akhiezer,Feinberg}.

\begin{figure}[tbh]
\begin{center}
\includegraphics[width=0.35\textwidth]{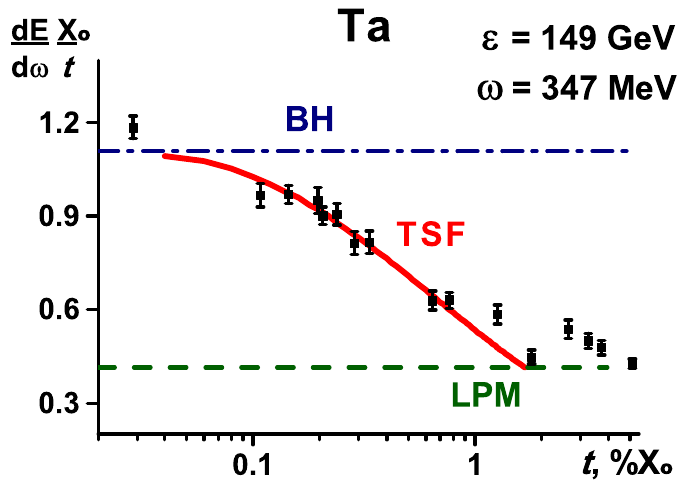}

\vspace{.4cm}

\includegraphics[width=0.35\textwidth]{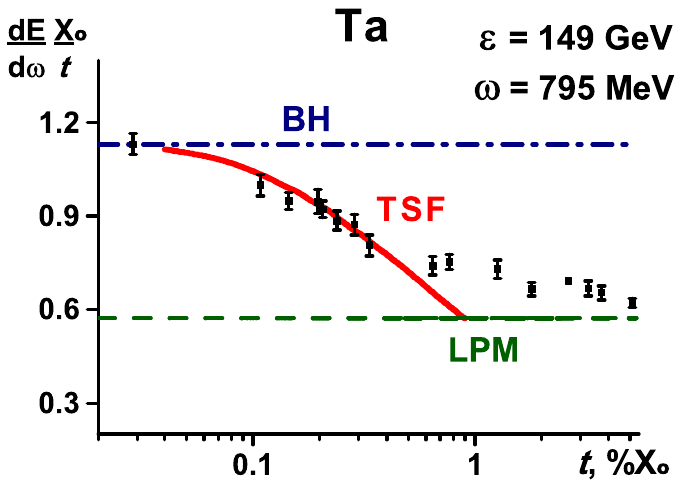}
\end{center}
\caption{The radiation power per unit length for $149\,$GeV electron radiation in tantalum target via target thickness $t$ $(\%\,X_{0})$. The detailed description of curves is given in text}
\label{fig:RadiationPerLength}
\end{figure}

The comparison of experimental data with the results of calculations using different approaches represented in
\cite{Thomsen:2009,Thomsen:2010}
shows a good not only qualitative, but also quantitative agreement.
In this short paper we present the comparison of the results of our calculation with the experimental data
\cite{Thomsen:2010}
only for two values of the emitted photon energy $\omega=347$ and $795\,$MeV (see Fig.~\ref{fig:RadiationPerLength}.
Following
\cite{Thomsen:2010}
we present here the radiation power spectrum per unit length, i.e. as $dE/d\omega$ multiplied by $X_{0}/t$. In these units the linear dependence of the radiation power spectrum for the BH (dot-dashed line) and LPM (dashed line) regimes of radiation are the constants (see Fig.~\ref{fig:RadiationPerLength}.
The curves TSF shows the logarithmic behavior of the radiation power spectrum in the intermediate region $l_{\rm c}>t>l_{\gamma}$ for given $\omega$.

For numerical calculations we used the original Fortran code based on the same formulas as the calculations of the TSF curves presented in figures in
\cite{Thomsen:2009,Thomsen:2010}.
Following
\cite{Thomsen:2010}
we took into account the multiphoton effect by corresponding normalization on the BH radiation spectrum.
The results of our calculations give a little excess (about $10\,\%$) over the results presented in
\cite{Thomsen:2009,Thomsen:2010}
by TSF curves in all figures, thereby they have a good agreement with experimental data (see, for example, Fig.~\ref{fig:RadiationPerLength}.
The essential discrepancy observing around the point $t=l_{\rm c}$ is easily explainable by the fact that the Migdal theory of the LPM effect is applicable at $t\gg{l_{\rm c}}$ while the Eq.~(\ref{eq:TSFspectra})
for the TSF regime of radiation is derived for $t\gg{l_{\rm c}}$. In the intermediate region $(2l_{\rm c}>t>l_{\rm c}/2)$ we have a smooth transition between these two regimes.

%
\section{\label{sec:angular}Angular distribution and polarization of radiation at non-dipole regime in crystal}

As it was shown in
\cite{Fomin:2003}
the non-dipole regime of radiation changes essentially not only spectrum of emitted gamma quanta, but also their angular distribution.
In
\cite{Fomin:2005}
it was proposed to use special features of angular characteristics of non-dipole coherent radiation in a thin crystal for production of the intensive photon beams with high degree of linear polarization.

This idea is based on the fact that the non-dipole regime of radiation, when the scattering angle becomes bigger than the characteristic angle of radiation of relativistic electron $\gamma^{-1}$, gives a possibility to avoid a mixture of the radiation emitted under the different (more than $\gamma^{-1}$) angles.
Using the photon collimators with angular width about $\gamma^{-1}$ one can organize the space-angular separation of photons emitted by electrons that were scattered at essentially different directions, for example, perpendicular.
To realize the non-dipole regime of radiation a high energy of electron beam is necessary.
However, it means that very narrow photon collimators should be used for this purpose: for the electron energy $\varepsilon=150\,$GeV the angle width of the collimator should be about $3\,\mu$rad. So, it is necessary to find the compromise condition for realization of this idea.

\begin{figure}[tbh]
\begin{center}
\includegraphics[width=0.5\textwidth]{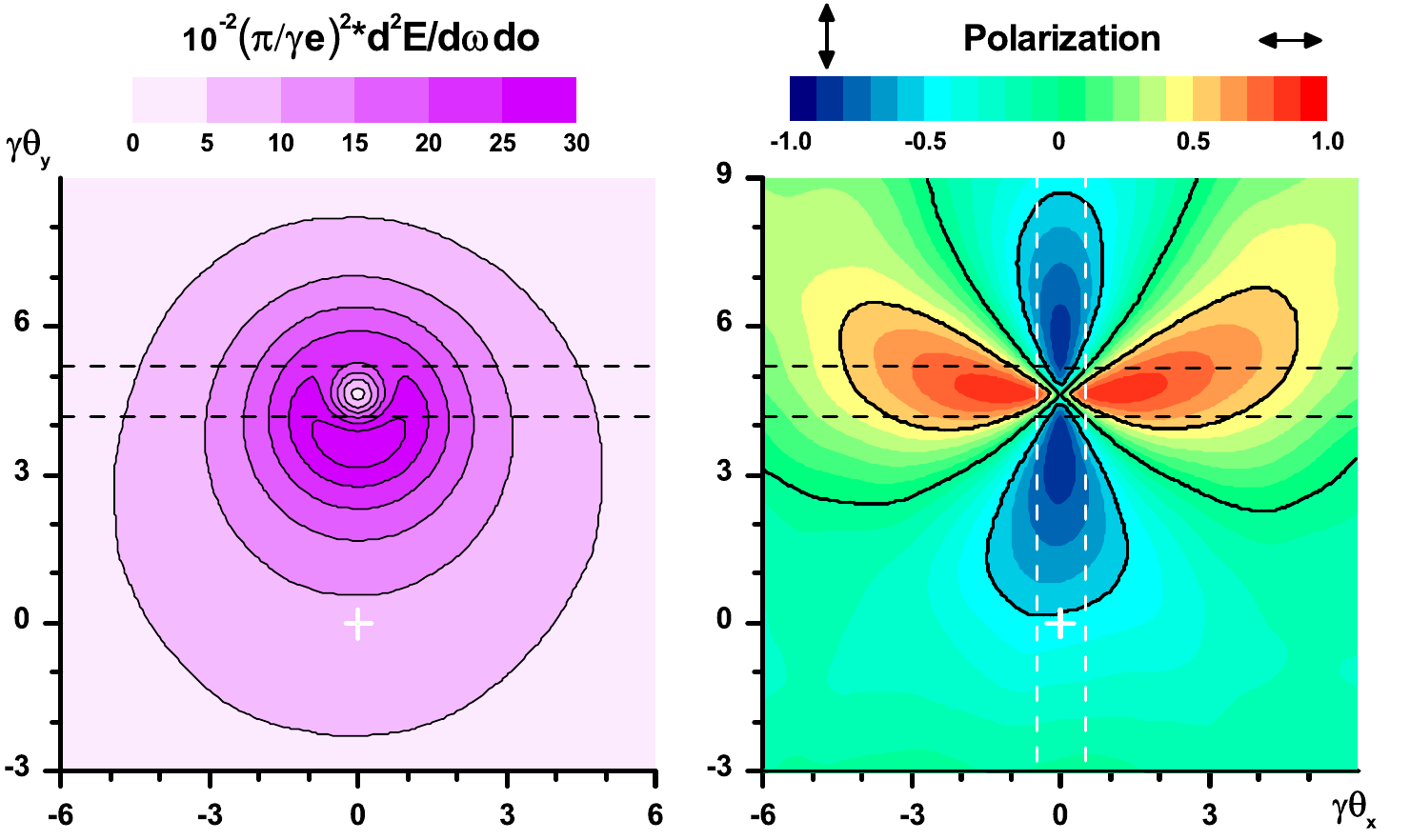}
\end{center}
\caption{The angular distributions (in units
$\gamma^{-1}$) of radiation power spectrum emitted by $3.5\,$GeV
electron beam incident on the tungsten monocrystal of $10\,\mu$m
thick at the angle $\psi_{\rm L}$ to the axis $<$111$>$ (left) and the
degree of linear polarization of this radiation (right)}
\label{fig:angular}
\end{figure}

To decrease the minimal electron energy for the non-dipole regime radiation is possible by using the coherent effect at relativistic electron scattering on atomic chains along the crystallographic axis (so called ``doughnut scattering effect", see i.e.
\cite{Akhiezer}).
The mean-square angle of multiple scattering in this case can exceed essentially the analogous parameter for amorphous matter
\cite{Shul'ga:1982}.
This effect is as strong, as high nuclear charge of the crystal material is, so, a good candidate for the crystal converter would be a tungsten monocrystal.

On the basis of the theoretical approach explained in details in
\cite{Fomin:2005}
we have carried out the calculations of the angular distributions and polarization of radiation by $3.5\,$GeV electrons incident on a tungsten crystal at the angle $\psi= \psi_{\rm L}$ to the axis $<$111$>$ (where $\psi_{\rm L}$ is the Lindhard angle
\cite{Lindhard}).
In this case $\theta_{\rm ms}\approx\psi_{\rm L}\,$=$\,0.6\,\mu$rad and the parameter of non-dipolity is $\gamma\theta_{\rm ms}$$\,\approx\,$4.
The results of 
the computer simulation (using binary collision model
\cite{Fomin:2005})
for $3.5\,$GeV electron scattering by the $10\,\mu$m tungsten crystal when the electron beam falls at the angle $\psi=\psi_{\rm L}$ to the crystal axis $<$111$>$
are presented in Fig.~\ref{fig:angular}.
The angular distribution of corresponding spectral-angular radiation density $d^{2}E/d\omega{do}$ of emitted photons is presented on the left part. The right part represents the angular distribution of the linear polarization degree of emitted photons from the $100\%$ vertically polarized photons $(P=-1)$ to $100\%$ horizontal polarization $(P=1)$.
All angles in Fig.~\ref{fig:angular}
are measured in units $\gamma^{-1}$.
The integral (over all angles) degree of linear polarization of radiation is close to zero.
However, using the slit-type horizontal (or vertical) photon collimator with the angular width $\Delta\theta_{\gamma}=\gamma^{-1}$ and putting it as shown in Fig.~\ref{fig:angular}
by dashed lines it is possible to obtain a linearly polarized (along the collimator plane) photon beam with polarization degree of about $60\,\%$.
Note, that the radiation intensity in the case of axially oriented crystal is much higher than at planar orientation case, which is applied normally for production of polarized photon beams.

%
\section{\label{sec:conclusion}Conclusion}

The comparison of the results of recent CERN experiment NA63 on measurement of radiation power spectrum of $149\,$GeV electrons in a series of thin tantalum targets
\cite{Thomsen:2010}
shows a good agreement with the corresponding calculations based on the quantitative theory of relativistic electron radiation in a thin layer of matter developed earlier in
\cite{Shul'ga:1998:Lett,Shul'ga:1998}.
The experimental observation of logarithmic dependence of radiation yield from the target thickness
\cite{Thomsen:2010}
is the first direct demonstration of the suppression of radiation effect for a relativistic electron with non-equilibrium own Coulomb field (TSF effect), which was predicted and theoretically studied in
\cite{Ternovskii,Shul'ga:1978,Fomin:1986}.
Note, that this effect should have its analog also in QCD at quark-gluon interaction.
The special features of angular distribution of radiation and its polarization in a thin target at non-dipole regime of radiation are proposed for a new experimental study.
These features can give a new opportunity for obtaining a high degree of linear polarization of gamma-quanta.

%
\section{\label{sec:acknowledgments}Acknowledgments}

We are very grateful to all participants of CERN collaboration NA63 for a brilliant performance of very complicated measurements of radiation in a set of ultra thin targets.
It makes enabled the first observation of the logarithmic thickness dependence of radiation power of relativistic electron at non-dipole regime that was predicted many years ago. Special thanks to Ulrik Uggerhoj for fruitful discussions of the subject of this investigation.

%
%

\end{document}